\documentclass[aip,jap,reprint,amsmath,amssymb,showpacs,showkeys,superscriptaddress]{revtex4-1}

\usepackage{graphicx}
\usepackage{dcolumn}
\usepackage{bm}
\usepackage[mathlines]{lineno}

\begin{document}


\title{Homogeneous limit of Cd$_{1\textrm{-}x}$Mn$_{x}$GeAs$_{2}$ alloy: electrical and magnetic properties}

\author{L.~Kilanski}
 \email{kilan@ifpan.edu.pl}
\author{M.~G\'{o}rska}
\author{E.~Dynowska}
\author{A.~Podg\'{o}rni}
\author{A.~Avdonin}
\author{W.~Dobrowolski}
\affiliation{Institute of Physics, Polish Academy of Sciences, Al. Lotnikow 32/46, 02-668 Warsaw, Poland}

\author{I.~V.~Fedorchenko}
\author{S.~F.~Marenkin}
\affiliation{Kurnakov Institute of General and Inorganic Chemistry RAS, 119991 Moscow, Russia}

\date{\today}

\begin{abstract}

We present the studies of structural, electrical, and magnetic properties of bulk Cd$_{1\textrm{-}x}$Mn$_{x}$GeAs$_{2}$ crystals with low Mn content, $x$, varying from 0 to 0.037. The studied samples have excellent crystallographic quality indicated by the presence of diffraction patterns never before observed experimentally for this compound. The electrical transport in our samples is dominated by thermal activation of conducting holes from the impurity states to the valence band with activation energy of about 200$\;$meV. The defect states acting as ionic scattering centers with concentration in the range from 6 to 15$\times$10$^{17}$$\;$cm$^{-3}$ are observed. The effective Mn content in our samples, $\bar{x}_{\theta}$, determined from fit of the susceptibility data to the Curie-Weiss law, is very close to the average chemical content, $x$. It indicates that the Mn ions are distributed randomly, substituting the Cd sites in the host CdGeAs$_{2}$ lattice. We observe a negative Curie-Weiss temperature, $|\theta|$$\,$$\leq$$\,$3.1$\;$K, increasing as a function of $x$. This indicates the significance of the short-range interactions between the Mn ions.

\end{abstract}

\keywords{semimagnetic-semiconductors; magnetic-impurity-interactions, exchange-interactions}

\pacs{72.80.Ga, 75.30.Hx, 75.30.Et, 75.50.Pp}



\maketitle

\section{Introduction}

The studies on semiconductors based on Mn-alloyed II-VI materials\cite{Kossut1993a, Dobrowolski2003a} have led to development of a new class of semiconductors called diluted magnetic semiconductors (DMS). The discovery of carrier mediated ferromagnetism in $p$-type IV-VI semiconductors (Ref.$\;$\onlinecite{Story1986a}) followed with development of III-V group compounds (Ref.$\;$\onlinecite{Ohno1996a}) resulted in considerable development of these materials in view of their possible application in spintronics. In DMS it is possible to study and independently tune and control their electronic and magnetic properties. Apart from classical III-V, IV-VI, and II-VI DMS, in which the Curie-Weiss temperature, $\theta$, does not exceed 200$\;$K, there are several other groups of prospective and intensively studied materials.\cite{Dietl2010a} \\ \indent Complex diluted magnetic semiconductors, such as II-IV-V$_{2}$ chalcopyrite materials, are perceived as prospective candidates for being used in spintronics.\cite{Erwin2004a, Picozzi2004a} Room temperature ferromagnetism in Zn$_{1\textrm{-}x}$Mn$_{x}$GeAs$_{2}$ and Cd$_{1\textrm{-}x}$Mn$_{x}$GeAs$_{2}$ alloys with $\theta$ as high as 367$\;$K for Zn$_{1\textrm{-}x}$Mn$_{x}$GeAs$_{2}$ with 3 at.\% of Mn\cite{Novotortsev2007a} was found (via direct observation of NMR spectra characteristic of MnAs hyperfine structure) to be related to the presence of MnAs clusters.\cite{Kilanski2008a, Kilanski2010a, Kilanski2011a} The combination of the semiconductor matrix and metallic clusters called nanocomposite can be prospective from the point of view of applicable magnetotransport effects. The MnAs clusters, however, interact with the conducting carriers inducing magnetoresistive effects with either negative values up to -50\% for Zn$_{1\textrm{-}x}$Mn$_{x}$GeAs$_{2}$ at $T$$\,$$=$$\,$1.4$\;$K (Refs.$\;$\onlinecite{Kilanski2010a} and \onlinecite{Kilanski2011b}) or positive values for Cd$_{1\textrm{-}x}$Mn$_{x}$GeAs$_{2}$ with 3 at. \% of Mn at $T$$\,$$=$$\,$1.4$\;$K (Ref.$\;$\onlinecite{Kilanski2011a}). However, in order to understand the complex magnetic properties of ferromagnetic semiconductor systems, it is necessary to study low paramagnetic ion alloying regime, where the aggregation of magnetic impurities does not occur. \\ \indent In the present paper we focused on the Cd$_{1\textrm{-}x}$Mn$_{x}$GeAs$_{2}$ crystals without any signatures of MnAs clusters, observed before in the samples with $x$$\,$$>$$\,$0.05.\cite{Kilanski2011a} The homogeneous Mn distribution in the Cd$_{1\textrm{-}x}$Mn$_{x}$GeAs$_{2}$ lattice can be accomplished by limiting the Mn content, $x$, to 0.037. The thermally activated transport was observed in our samples with impurity states with activation energy $E_{a}$$\,$$=$$\,$200$\;$meV. Our samples showed paramagnetism with the effective Mn-concentration reduced with respect to the average Mn content, $x$.

\section{Structural characterization}

\begin{figure}
 \includegraphics[width = 0.42\textwidth, bb = 0 20 590 580]{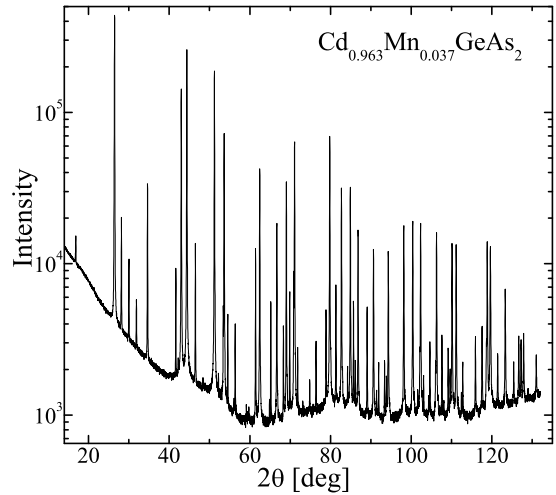}%
 \caption{\label{FigXRD} Diffraction patterns observed for the exemplary Cd$_{1\textrm{-}x}$Mn$_{x}$GeAs$_{2}$ sample with $x$$\,$$=$$\,$0.037.}
 \end{figure}

For purpose of this research bulk Cd$_{1\textrm{-}x}$Mn$_{x}$GeAs$_{2}$ crystals with low Mn content, $x$, varying from 0 up to 0.037, were prepared by using the vertical Bridgman method.\cite{Marenkin1999a} Our samples were synthesized from stoichiometric ratios of CdAs$_{2}$, Ge, and Mn of at least 5N purity. The sample synthesis was done in  vacuum sealed (to a pressure of about 10$^{-2}$$\;$Pa) graphitized quartz ampoules with the CdGeAs$_{2}$ seed. The growth was performed at the temperature of about 950$\pm$0.5$\;$K. The growth was finished with rapid cooling (about 5-10 K/s) down to room temperature in order to prevent segregation of magnetic impurities and increase the uniformity of the as-grown crystals. A more detailed description of the growth procedure can be found in Refs.$\;$\onlinecite{Marenkin2004a, Marenkin2005a}. \\ \indent The chemical composition of the as-grown samples was determined using the energy dispersive x-ray fluorescence method (EDXRF). Measurements were done at room temperature with the use of the Tracor X-ray Spectrace 5000 Spectrometer equipped with Si(Li) detector. The as-grown Cd$_{1\textrm{-}x}$Mn$_{x}$GeAs$_{2}$ ingots were cut into 1$\;$mm-thick slices perpendicular to the growth direction. The measured Mn molar fraction, $x$, determined with maximum relative uncertainty of EDXRF method not exceeding 10\%, is found to be almost constant as a function of the ingot length. The EDXRF data indicate that all our samples have a proper stoichiometry of the compound equal to 1:1:2. The EDXRF spectra show no evidence of the fluorescent lines coming from unwanted dopants in the studied alloy. This means that unintended impurity concentrations in the alloy are less than 10$^{15}$$\;$cm$^{-3}$. The values of the chemical composition, $x$, together with the estimated experimental errors are gathered in Table$\;$\ref{tab:BasicCharact}. The EDXRF analysis indicates that our samples have chemical composition, $x$, changing from 0 for pure CdGeAs$_{2}$ crystal up to 0.037. The present samples are of a major improvement with respect to our older crystals (see Ref.$\;$\onlinecite{Kilanski2011a}) due to good stoichiometry. \\ \indent The structural quality of our samples was studied with the use of the high resolution x-ray diffraction method (HRXRD). We used multipurpose diffractometer (X'Pert PRO MPD, Panalytical configured for Bragg-Brentano diffraction geometry) equipped with a strip detector and an incident-beam Johansson monochromator. The Cu K$_{\alpha 1}$ x-ray radiation with wavelength equal to 1.5406$\textrm{\AA}$ was used. This instrument allows obtaining diffraction patterns with an excellent resolution and counting statistics. The obtained HRXRD patterns for both pure CdGeAs$_{2}$ crystal and Mn-alloyed samples show excellent crystalline quality of our samples (see Fig.$\;$\ref{FigXRD}).
The crystal structure of our samples is improved with respect to the old results.\cite{Marenkin2004a, Novotortsev2005a} The HRXRD patterns obtained for all our samples were analyzed with the use of Rietveld refinement method. The data analysis indicates that all our samples crystallized in tetragonal chalcopyrite structure. The Rietveld fits allowed precise calculation of the lattice parameters for the samples with different chemical composition, $x$. The results of our calculations are presented in Table$\;$\ref{tab:BasicCharact}.
\begin{table}[t]
\caption{\label{tab:BasicCharact}
Preliminary results of a basic structural characterization including the chemical composition, $x$, and the chalcopyrite structure lattice parameters, $a$ and $c$, respectively.}
\begin{tabular}{ccc}
\hline  \hline
 $x$$\,$$\pm$$\,$$\Delta$$x$ & $a$$\,$$\pm$$\,$$\Delta$$a$ (\AA) & $c$$\,$$\pm$$\,$$\Delta$$c$ (\AA)  \\ \hline
 0                       & 5.9452$\,$$\pm$$\,$0.0002 & 11.2153$\,$$\pm$$\,$0.0003   \\
 0.004$\,$$\pm$$\,$0.001 & 5.9425$\,$$\pm$$\,$0.0002 & 11.2207$\,$$\pm$$\,$0.0006    \\
 0.013$\,$$\pm$$\,$0.001 & 5.9439$\,$$\pm$$\,$0.0002 & 11.2156$\,$$\pm$$\,$0.0006    \\
 0.024$\,$$\pm$$\,$0.002 & 5.9436$\,$$\pm$$\,$0.0002 & 11.2116$\,$$\pm$$\,$0.0004    \\
 0.037$\,$$\pm$$\,$0.003 & 5.9417$\,$$\pm$$\,$0.0003 & 11.2082$\,$$\pm$$\,$0.0006    \\ \hline  \hline
\end{tabular}
\end{table}
The lattice parameters are Mn composition dependent. The $a$($x$) and $c$($x$) dependencies are decreasing functions of the Mn content for most of our crystals with exception of the sample with $x$$\,$$=$$\,$0.004. However, the volume of the unit-cell $V$ is a monotonic, decreasing function of $x$. It indicates that in the case of the two samples with the smallest Mn content in our series, the deformation of the lattice constant occurred along the different crystallographic axes. The lattice parameters obtained by us for pure CdGeAs$_{2}$ are close to the literature data for this compound\cite{Pfister1958a}: $a$$\,$$=$$\,$5.942$\;${\AA} and $c$$\,$$=$$\,$11.224$\;${\AA}. Nearly monotonic, decreasing $a$($x$) and $c$($x$) dependencies suggest that the Mn ions substitute the Cd ions in the crystal lattice, since the tetrahedral radius of the Mn ion is smaller than that of Cd.\cite{Pyykko2012a} The $a$($x$) and $c$($x$) dependencies are fitted with the linear function of $x$ with good agreement between the experimental data and the fitted lines. The lattice parameters change as a function of $x$ with a slope similar to that presented in Ref.$\;$\onlinecite{Novotortsev2005a}. A more detailed description of the HRXRD data can be found in Ref.$\;$\onlinecite{Dynowska2013a}.

\section{Magnetotransport data}

\begin{figure}
 \includegraphics[width = 0.42\textwidth, bb = 0 20 590 580]{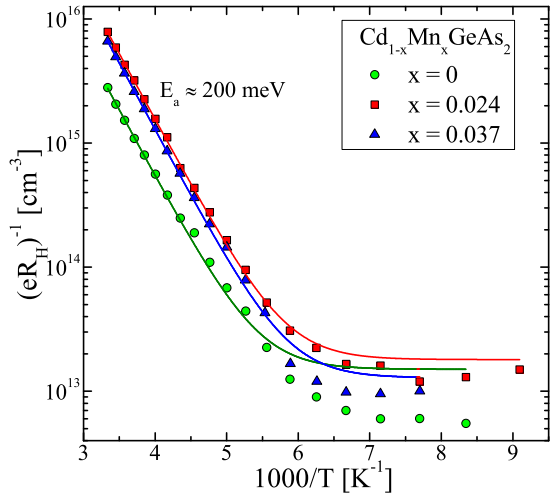}%
 \caption{\label{FignvsT} The Hall carrier concentration $p$$\,$$=$$\,$($e R_{H}$)$^{-1}$ as a function of the inverse of the temperature for the selected Cd$_{1\textrm{-}x}$Mn$_{x}$GeAs$_{2}$ samples with different chemical compositions (symbols). The lines show fits to the thermally activated carrier transport model.}
 \end{figure}

\begin{figure}
 \includegraphics[width = 0.42\textwidth, bb = 0 20 590 580]{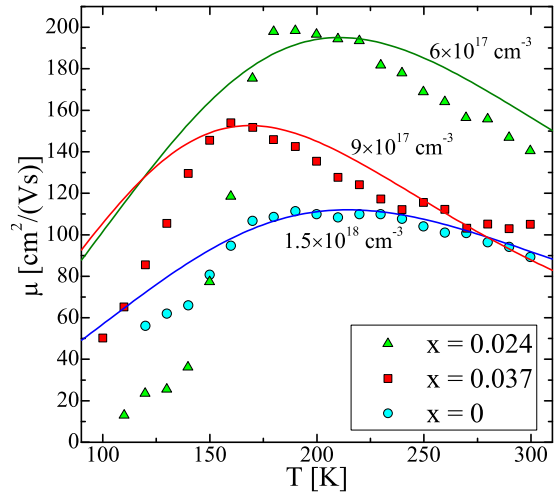}%
 \caption{\label{FigmuvsT} The Hall carrier mobility as a function of the temperature for selected Cd$_{1\textrm{-}x}$Mn$_{x}$GeAs$_{2}$ samples with different chemical compositions. The markers represent the experimental data and the lines are obtained form fits to Eq.$\;$\ref{EqEmpMobGaAs}. The fitted concentrations, $N_{I}$, are shown next to the lines.}
 \end{figure}

In order to obtain fundamental information about the electrical transport in Cd$_{1\textrm{-}x}$Mn$_{x}$GeAs$_{2}$ crystals with varying chemical composition we performed the temperature dependent Hall effect measurements. The Hall effect measurements were done using a standard six contact dc current technique in the presence of static magnetic field of induction not exceeding $B$$\,$$=$$\,$13$\;$T. Prior to magnetotransport measurements the samples were cut into Hall bars with dimensions of about 1$\times$1$\times$8$\;$mm$^{3}$, chemically cleaned and etched. The contacts to the samples were made using gold wire and indium solder. The measurements were done over a broad temperature range from $T$$\,$$=$$\,$1.4$\;$K up to 300$\;$K. \\ \indent We measured a series of isothermal magnetic field dependencies of the off-diagonal resistivity component $\rho_{xy}$($B$). The obtained $\rho_{xy}$($B$) dependencies are linear over the studied temperature range. No anomalous Hall effect was observed in our samples. As a result of the $\rho_{xy}$($B$) measurements we calculated the temperature dependence of the Hall carrier concentration $p$ for all our Cd$_{1\textrm{-}x}$Mn$_{x}$GeAs$_{2}$ samples presented in Fig.$\;$\ref{FignvsT}.
Our samples became highly resistive at temperatures lower than 100$\;$K and the Hall constant became unmeasurable. The Hall constant for all our samples points into the $p$-type conductivity with relatively low carrier concentration, $p$$\,$$<$$\,$10$^{16}$$\;$cm$^{-3}$, at room temperature.  The Hall effect results show that the carrier transport in our samples is thermally activated in the valence band at temperatures higher than 150$\;$K. The data gathered in Fig.$\;$\ref{FignvsT} allowed us to estimate the activation energy of band carriers, $E_{a}$. The calculated values of $E_{a}$ are almost constant as a function of the Mn-content, $x$, and are equal to about 200$\;$meV. Similar activation energies for both pure and Mn-alloyed samples together with the fact that Mn is isovalent in chalcopyrite structure indicates that the Mn ions are not directly related to the presence of impurity states.  It is then evident that other defect types, possibly antisite defects\cite{Paudel2008a} that are the main defect type in CdGeAs$_{2}$, because they have lower formation energies than Cd or Ge vacancy defects, are present with moderate concentration in our samples and are responsible for the observed impurity states. \\ \indent The Hall carrier mobility, $\mu$, is defined from the equation $\sigma_{xx}$$\,$$=$$\,$$e p \mu$, where $\sigma_{xx}$ is the conductivity tensor component parallel to the current direction, measured in the absence of the external magnetic field, and $e$ is the elementary charge. The carrier mobility as a function of the temperature, calculated for each studied sample, is presented for selected Cd$_{1\textrm{-}x}$Mn$_{x}$GeAs$_{2}$ crystals with different chemical compositions in Fig.$\;$\ref{FigmuvsT}.
The data indicate that in all our samples above $T$$\,$$\approx$$\,$180$\;$K the phonon scattering is responsible for the decrease of the carrier mobility with increasing the temperature. The maximum value of the carrier mobility for our samples is not proportional to the average Mn content, $x$. It indicates, that Mn ions are not directly responsible for the carrier scattering in our samples. This conclusion is in agreement with the data gathered in Fig.$\;$\ref{FignvsT}. Below $T$$\,$$\approx$$\,$180$\;$K a decrease of the mobility with decreasing the temperature is connected with impurity scattering. The $\mu$($T$) dependence at temperatures between 100$\;$K and 170$\;$K can be fairly well fitted to the $T^{3/2}$ dependence, indicating that the concentration of scattering centers is temperature independent due to compensation. \\ \indent The Hall carrier mobility in $p$-type GaAs can be fitted with the use of empirical two-band model in the low-field approximation, describing the hole conductivity in valence and impurity bands. The Hall effect scattering factors are assumed to be equal to unity for all our samples. The empirical mobility, as described in Ref.$\;$\onlinecite{Blakemore1982a}, may be fitted to the expression
\begin{equation}\label{EqEmpMobGaAs}
    \frac{1}{\mu} = \frac{1}{A \cdot T^{3/2} \cdot N_{I}^{-1} } + \frac{1}{B \cdot T^{-2.3}},
\end{equation}
where $N_{I}$ is the concentration of scattering centers, $A$ and $B$ are constants describing the ionized impurity scattering and the combined scattering due to phonons, respectively. For our purpose we took the $A$ and $B$ values from GaAs, a binary equivalent of ternary CdGeAs$_{2}$. At low temperatures, where ionic impurity scattering processes are of a major importance the $A$ value is equal to 2.5$\times$10$^{20}$$\;$(cm$\cdot$V$\cdot$s)$^{-1}$, while at high temperatures, where phonon scattering is the main scattering mechanism, the value of $B$ equals 2.5$\times$10$^{3}$.\cite{Blakemore1982a} We performed fitting of the experimental $\mu$($T$) data (symbols in Fig.$\;$\ref{FigmuvsT}) to the Eq.$\;$\ref{EqEmpMobGaAs} with $A$$\,$$=$$\,$2.5$\times$10$^{20}$$\;$(cm$\cdot$V$\cdot$s)$^{-1}$, $B$$\,$$=$$\,$2.5$\times$10$^{3}$, and $N_{I}$ as the only fitting parameter. As a result of the fitting procedure the $N_{I}$ values were estimated for each sample. The concentration of the ionic scattering centers, $N_{I}$, changes from 6$\times$10$^{17}$$\;$cm$^{-3}$ to 1.5$\times$10$^{18}$$\;$cm$^{-3}$, for the sample with $x$$\,$$=$$\,$0.024 and 0, respectively. The obtained defect concentration values for the samples with different $x$ are not proportional to the amount of Mn. It indicates that the Mn incorporation in the CdGeAs$_{2}$ does not produce ionic scattering centers. \\ \indent The magnetoresistance measurements were also done in parallel with the Hall effect measurements. All our samples have high resistance and an absence of free carriers at temperatures lower than 100$\;$K. At $T$$\,$$>$$\,$100$\;$K the obtained magnetoresistance curves for all our samples indicate the presence of positive magnetoresistance with relative amplitudes not exceeding 2\%. The positive magnetoresistance scales with the square of the magnetic field indicating that this effect is the classical effect due to the orbital carrier movement in the presence of an applied magnetic field.

\section{Magnetic properties}

\begin{figure}
 \includegraphics[width = 0.42\textwidth, bb = 0 20 590 560]{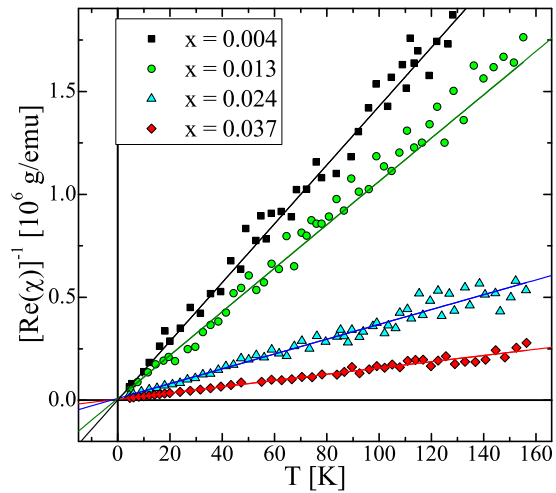}%
 \caption{\label{FignXsT} The inverse of the real part of the dynamic magnetic susceptibility measured (symbols) for the selected Cd$_{1\textrm{-}x}$Mn$_{x}$GeAs$_{2}$ samples with different chemical composition. The lines represent theoretical curves fitted to the Eq.$\;$\ref{EqCWLaw}}.
 \end{figure}

\begin{figure}
 \includegraphics[width = 0.42\textwidth, bb = 0 20 590 580]{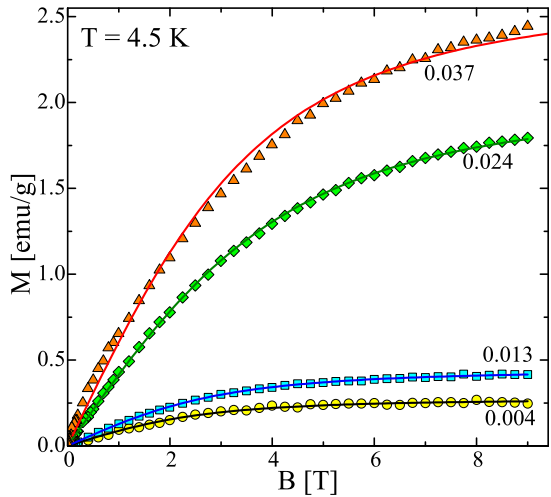}%
 \caption{\label{FignMvsB} The magnetization as a function of the applied magnetic field measured (symbols) at $T$$\,$$=$$\,$4.5$\;$K for the selected  Cd$_{1\textrm{-}x}$Mn$_{x}$GeAs$_{2}$ samples with different chemical compositions. The lines represent the theoretical curves fitted to the Eq.$\;$\ref{EqMSBrill}.}
 \end{figure}

\noindent Magnetic properties of our Cd$_{1\textrm{-}x}$Mn$_{x}$GeAs$_{2}$ samples were studied with the use of LakeShore  magnetometer/susceptometer system. This instrument allows measurements of both static and dynamic magnetic properties of solids. The temperature dependent dynamic magnetic susceptibility as well as the magnetic field dependencies were measured. \\ \indent The dynamic magnetic susceptibility, $\chi$, was measured over the temperature range from 4.3 up to 200 K. The sample was put into the alternating magnetic field with frequency $f$ equal to 625$\;$Hz and amplitude $H_{AC}$$\,$$=$$\,$10$\;$Oe. As a result, we obtained the temperature dependencies of both real and imaginary parts of the ac susceptibility. The measurements performed on all our Cd$_{1\textrm{-}x}$Mn$_{x}$GeAs$_{2}$ crystals indicated a vanishing imaginary part of the ac magnetic susceptibility. The temperature dependence of the inverse of the real part of the ac magnetic susceptibility for the selected Cd$_{1\textrm{-}x}$Mn$_{x}$GeAs$_{2}$ samples is presented in Fig.$\;$\ref{FignXsT}.
Our results clearly indicate the paramagnetic behavior of the (Re($\chi_{AC}$))$^{-1}$($T$) for all our samples. The temperature dependence of the inverse of the magnetic susceptibility, (Re($\chi_{AC}$))$^{-1}$($T$), was fitted over the temperature range 20 $-$ 160$\;$K to the Curie-Weiss expression of the form
\begin{equation}\label{EqCWLaw}
    \chi(T) = \frac{C}{T - \theta} + \chi_{dia},
\end{equation}
where
\begin{equation}\label{EqCWLawCConst}
    C= \frac{N_{0} g^{2} \mu_{B}^{2} S(S+1) {\bar x_{\theta}}}{3k_{B}}.
\end{equation}
Here $C$ is the Curie constant, $\chi_{dia}$$\,$$=$$\,$$-2.5\times$10$^{-7}$$\;$emu/g is the diamagnetic contribution to the magnetic susceptibility originating from the host lattice (the value was determined from our magnetization measurements of CdGeAs$_{2}$), $N_{0}$ is the number of cation sites per gram, $g$ is the g-factor of the magnetic ion (for Mn $g$$\,$$=$$\,$2), $S$$\,$$=$$\,$5/2 is the spin-magnetic momentum of the Mn ions, $\mu_{B}$ is the Bohr magneton, $k_{B}$ is the Boltzmann constant, and ${\bar x_{\theta}}$ is the effective magnetically-active Mn content. \\ \indent The experimental data gathered in Fig.$\;$\ref{FignXsT} were fitted to the Eq.$\;$\ref{EqCWLaw} assuming the constant value of the diamagnetic contribution to the magnetic susceptibility $\chi_{dia}$$\,$$=$$\,$$-2.5\times$10$^{-7}$$\;$emu/g estimated above for the pure CdGeAs$_{2}$ sample. We fitted the experimental (Re($\chi_{AC}$))$^{-1}$($T$) curves with two fitting parameters: the Curie-Weiss temperature $\theta$ and the Curie constant $C$. The fitted curves are presented together with the experimental data in Fig.$\;$\ref{FignXsT}. As we can clearly see the magnetic susceptibility of our samples can be very well reproduced with the use of the Curie-Weiss law defined with Eq.$\;$\ref{EqCWLaw}. The fitting parameters, $\theta$ and $C$, are gathered for all our samples in Table$\;$\ref{tab:MagnParams1}.
\begin{table}[t]
\caption{\label{tab:MagnParams1}
Susceptibility parameters for Cd$_{1\textrm{-}x}$Mn$_{x}$GeAs$_{2}$.}
\begin{tabular}{ccccccc}
\hline \hline
 $x$ &  C (10$^{-5}$) & $\bar{x}_{\theta}$ &  $\theta$ [K] & $J/k_{B}$ [K]  \\
 &  [emu$\cdot$K/g] & &  &   &  \\ \hline
 0.004  &  7.0$\pm$0.2  & 0.005$\pm$0.001 & -0.08$\pm$0.02 & -0.46    \\
 0.013  &  9.4$\pm$0.2  & 0.007$\pm$0.001 & -0.41$\pm$0.03 & -1.69        \\
 0.024  &   28$\pm$2    & 0.021$\pm$0.002 & -2.4$\pm$0.2   & -3.26      \\
 0.037  &   66$\pm$3    & 0.050$\pm$0.005 & -3.1$\pm$0.3   & -1.77        \\ \hline \hline
\end{tabular}
\end{table}
The Curie-Weiss temperatures, determined for all our samples are negative with the values increasing as a function of the Mn content, $x$. This indicates the significance of the short-range interactions between the Mn ions. The Curie constant values obtained for our samples, gathered in Table$\;$\ref{tab:MagnParams1}, are an increasing function of the Mn content. It indicates an increase in the amount of magnetically active Mn ions with increasing $x$ in our samples. The values of the Curie constant can be used to calculate the effective Mn content, $\bar{x}_{\theta}$, using Eq.$\;$\ref{EqCWLawCConst}. The estimated $\bar{x}_{\theta}$ values are gathered in Table$\;$\ref{tab:MagnParams1}. The discrepancy between the $x$ and $\bar{x}_{\theta}$ values is almost zero (within the accuracy of the estimation of both quantities) for the sample with the smallest $x$$\,$$=$$\,$0.004. For higher $x$ the difference between $x$ and $\bar{x}_{\theta}$ increases. It is a signature that the distribution of the Mn ions in the Cd$_{1\textrm{-}x}$Mn$_{x}$GeAs$_{2}$ lattice becomes imperfect. There are two explanations of this difference: (i) the charge state of a fraction of Mn ions might be different than Mn$^{2+}$, reducing its total magnetic momentum and (ii) antiferromagnetic Mn pairing leading to turning off their magnetic moment. However, it should be emphasized, that no signatures of the presence of large MnAs or other clusters are present in our samples. For the highest Mn content, $x$$\,$$=$$\,$0.037, the $\bar{x}_{\theta}$ value is higher than $x$. The difference (taking into account the uncertainties in $x$ and $\bar{x}_{\theta}$ estimation) might originate from clustering and/or underestimation of the experimental errors. \\ \indent Gudenko \emph{et al.} studied the electron paramagnetic resonance (EPR) in Cd$_{1\textrm{-}x}$Mn$_{x}$GeAs$_{2}$ with $x$$\,$$=$$\,$0.06.\cite{Gudenko2005a} They reported two Curie-Weiss temperatures: $\theta_{1}$$\,$$=$$\,$255$\;$K and $\theta_{2}$$\,$$=$$\,$$-$3$\;$K. The Curie-Weiss temperatures for our samples are close to $\theta_{2}$. Gudenko \emph{et al.} attributed the $\theta_{2}$ to the centers where Mn ions substitute the Cd ions. We suppose that this is the case in all our samples. \\ \indent Since our experimental results show that $\theta$$\,$$\ll$$\,$$T$, we can use $\theta$ to estimate the exchange interaction, which we have assumed to be the nearest-neighbor exchange, from the relation
\begin{equation}\label{EqExchInter}
    J/k_{B} = \frac{3 \theta}{ \bar{x}_{\theta} S(S+1)z},
\end{equation}
where $z$ is the number of nearest-neighbor cation sites, 12 for the chalcopyrite structure. (We remember that in our chalcopyrite compound $x$ refers to one kind of cations, i.e., half of the total cation number.) \\ \indent In the last column of Table$\;$\ref{tab:MagnParams1} we present the values of the exchange parameter, $J/k_{B}$, for all our samples. The errors, estimated from the scatter in the experimental data and from the uncertainties in the fitting, are about $\pm$30\% for $J/k_{B}$. The values of $J/k_{B}$ are negative and relatively small, characteristic for the antiferromagnetic superexchange via an anion.\cite{Spalek1986a, Gorska1988a} They are of the same order but somewhat smaller than those for zinc-blende II-VI DMS with Mn\cite{Spalek1986a}.  Larson \emph{et al.} suggested that in the superexchange interaction the anion is of more importance than the nonmagnetic cation. \cite{Larson1985a} Therefore, it would be also interesting to compare our results with magnetic properties of compounds with group V anion, e.g., III-V DMS. This is not trivial, since in Mn-doped III-V DMS the divalent manganese substitutes the trivalent group III cation. That results in high hole concentration, usually above 10$^{21}$$\;$cm$^{-3}$. In materials with such high free carrier concentration we observe a carrier induced ferromagnetism due to the RKKY exchange interaction and ferromagnetic Curie-Weiss temperatures of an order of 100 K (for a review see e.g. Ref.$\;$\onlinecite{Dietl2010a}). To find III-V DMS with magnetic properties like in our material we must consider only very dilute III-V DMS in which the free carrier concentration does not exceed 10$^{19}$$\;$cm$^{-3}$.  In In$_{1\textrm{-}x}$Mn$_{x}$As with free carrier concentration below 10$^{19}$$\;$cm$^{-3}$ $J/k_{B}$ equals -1.6$\;$K (Ref.$\;$\onlinecite{Molnar1991a}) and in Ga$_{1\textrm{-}x}$Mn$_{x}$N with free carrier concentration below 10$^{18}$$\;$cm$^{-3}$ $J/k_{B}$ equals -1.9$\;$K(Ref.$\;$\onlinecite{Zajac2001a}). These values are very similar to the values of $ J/k_{B}$ observed here. Our results indicate, that the superexchange interaction in chalcopyrite DMS depends strongly on the type of the chemical bond, as was suggested for other DMS experimentally\cite{Spalek1986a, Gorska1988a} and theoretically.\cite{Larson1985a} \\ \indent The static magnetic properties of Cd$_{1\textrm{-}x}$Mn$_{x}$GeAs$_{2}$ samples were studied with the use of the Weiss extraction method employed to the LakeShore 7229 susceptometer/magnetometer system. This instrument allowed us to study the magnetic moment at different temperatures from 4.3 up to 200$\;$K and in the presence of an external magnetic field with maximum induction not exceeding $B$$\,$$=$$\,$9$\;$T. Our studies covered the isothermal measurements of the magnetization $M$ as a function of the magnetic field $B$ done at few selected temperatures $T$$\,$$<$$\,$200$\;$K. The magnetization as a function of the magnetic field measured at $T$$\,$$=$$\,$4.5$\;$K for the selected Cd$_{1\textrm{-}x}$Mn$_{x}$GeAs$_{2}$ crystals with different chemical composition, $x$, is presented in Fig.$\;$\ref{FignMvsB}.
The $M$($B$) measurements showed the lack of measurable magnetic irreversibility of our samples. The magnetization results obtained at $T$$\,$$>$$\,$5$\;$K are typical of paramagnetic materials. \\ \indent Because the $M$($B$) curves at low temperatures carry most information about the paramagnetic media and for clarity of presentation only the results obtained at the lowest measured temperature, $T$$\,$$=$$\,$4.3$\;$K, are presented and analyzed. \\ \indent The $M$($B$) curve for the nonmagnetic CdGeAs$_{2}$ sample showed a diamagnetic, negative slope. The diamagnetic susceptibility for CdGeAs$_{2}$ crystal, estimated from the slope of the $M$($B$) curve, $\chi_{dia}$$\,$$=$$\,$$-2.5\times$10$^{-7}$$\;$emu/g. This result is in agreement with the diamagnetic susceptibility value determined from the ac susceptibility data. The magnetization for the samples with $x$$\,$$\leq$$\,$0.013 reaches saturation in moderate magnetic fields, $B$$\,$$<$$\,$9$\;$T, used in our experiments. However, for $x$$\,$$>$$\,$0.013 the magnetization curves do not show saturation. Therefore, the $M$($B$) curves had to be fitted with the modified Brillouin function in order to properly estimate the saturation magnetization of our samples and the role of short range Mn-pairing. We fitted our experimental data to the expression\cite{Gaj1979a}
\begin{equation}\label{EqMSBrill}
    M = M_{S} B_{S}\Bigg{(}\frac{g \mu_{B} S B}{k_{B} (T + T_{0})}\Bigg{)}+ \chi_{dia} B,
\end{equation}
where
\begin{equation}\label{EqMSx}
    M_{S} = \bar{x}_{m} N_{0} \mu_{B} g S.
\end{equation}
Here $B_{S}$ is the Brillouin function and $\bar{x}_{m}$ is the effective, active Mn content estimated from the saturation magnetization values. The term $\chi_{dia} B$ represents the diamagnetic contribution of the CdGeAs$_{2}$. The experimental $M$($B$) curves were fitted with two fitting parameters: $M_{S}$ and $T_{0}$. The results of the fitting procedure are presented in Fig.$\;$\ref{FignMvsB} together with the experimental data. As we can see the experimental curves are well fitted with the theoretical model. The fitting parameters, obtained during the fitting procedure, are gathered for all our samples in Table$\;$\ref{tab:MagnParams2}.
\begin{table}[t]
\caption{\label{tab:MagnParams2}
Parameters for the modified Brillouin function fit.}
\begin{tabular}{cccc}
\hline \hline
 $x$    &  $M_{S}$ [emu/g] &  $\bar{x}_{m}$    &  $T_{0}$ [K]  \\ \hline
 0.004  &  0.27$\pm$0.02   &  0.005$\pm$0.001  & 0.061$\pm$0.012   \\
 0.013  &  0.43$\pm$0.03   &  0.008$\pm$0.002  & 0.63$\pm$0.12   \\
 0.024  &  2.0$\pm$0.2     &  0.037$\pm$0.005  & 2.8$\pm$0.5     \\
 0.037  &  2.7$\pm$0.3     &  0.050$\pm$0.008  & 2.0$\pm$0.4     \\ \hline \hline
\end{tabular}
\end{table}
The $T_{0}$ temperatures have positive values. Positive  $T_{0}$ in Eq.$\;$\ref{EqMSBrill} corresponds to negative  $\theta$ in Eq.$\;$\ref{EqCWLaw}. The values of $T_{0}$ change with the Mn content, $x$, in a way similar to the Curie-Weiss temperature, $\theta$, within the accuracy of estimation of both quantities. We estimate the error in $T_{0}$ to be about 20\%. The only exception occurs for the sample with $x$$\,$$=$$\,$0.037; that might originate from the increasing influence of Mn-pairs or short-range clustering in the magnetism of this compound. The saturation magnetization, $M_{S}$, estimated for our samples is an increasing function of $x$. The $M_{S}$ values were used in order to calculate the amount of magnetically active Mn ions, $\bar{x}_{m}$, using Eq.$\;$\ref{EqMSx}. For the samples with the low Mn content, $x$$\,$$\leq$$\,$0.013, the calculated $\bar{x}_{m}$ values (Table$\;$\ref{tab:MagnParams2}) are very close to $x$ and  $\bar{x}_{\theta}$. For higher $x$, however, we have $x$$\,$$\approx$$\,$$\bar{x}_{\theta}$$\,$$\leq$$\,$$\bar{x}_{m}$ for $x$$\,$$=$$\,$0.024 and $x$$\,$$\leq$$\,$$\bar{x}_{\theta}$$\,$$\approx$$\,$$\bar{x}_{m}$ for $x$$\,$$=$$\,$0.037. The discrepancy between the values, especially visible for $x$$\,$$=$$\,$0.037, might originate from local imperfections in the sample. We have very good agreement between the susceptibility and magnetization data. However, it is also possible that we underestimated the error in parameter estimation for the sample with $x$$\,$$=$$\,$0.037. It must be emphasized, that our data indicate a rather good distribution of Mn ions in our samples.

\section{Summary}

We explored the structural, electrical, and magnetic properties of bulk Cd$_{1\textrm{-}x}$Mn$_{x}$GeAs$_{2}$ crystals with low Mn content, $x$, varying from 0 to 0.037. Our samples have good structural quality, with lattice parameters changing as a function of the Mn content, $x$, according to the Vegard rule. \\ \indent The transport properties of our samples show $p$-type conductivity due to impurity states present near valence band, with activation energy of about 200$\;$meV. The carrier transport in our samples at temperatures higher than 150$\;$K is thermally activated, while at $T$$\,$$<$$\,$150 K a degenerated transport dominates. The carrier mobility shows the presence of ionic carrier scattering centers with concentration from 6 to 15$\times$10$^{17}$$\;$cm$^{-3}$. The scattering centers are not related to the Mn impurities. \\ \indent The paramagnetic Curie-Weiss behavior indicates that the majority of Mn ions in our samples are isolated (randomly distributed in the host CdGeAs$_{2}$ lattice). The estimates of the active Mn content of our samples, in case of most of them, give values similar to those obtained with the use of the EDXRF method. It indicates that the majority of Mn ions substitute the Cd positions in the CdGeAs$_{2}$ lattice, and possess Mn$^{2+}$ charge state with high magnetic momentum. We prove that the total solubility of Mn in Cd$_{1\textrm{-}x}$Mn$_{x}$GeAs$_{2}$ is around  $x$$\,$$=$$\,$0.04, a value lower than for II-VI and rather high with respect to III-V DMS bulk crystals grown under thermal equilibrium conditions. Both II-VI and III-V DMS we may consider as ternary analogs of Cd$_{1\textrm{-}x}$Mn$_{x}$GeAs$_{2}$. The negative Curie-Weiss temperatures, with values increasing as a function of $x$ are observed indicating an antiferromagnetic exchange interaction. The average value of the exchange parameter, $J/k_{B}$$\,$$\approx$$\,$-1.8$\pm$1$\;$K, is of the same order as those observed in II-VI DMS and in very dilute III-V DMS with carrier concentration below 10$^{19}$$\;$cm$^{-3}$. This result may be explained by assuming the superexchange via an anion.

\section{Acknowledgments}

\noindent Scientific work was financed from funds for science in 2011-2014, under the project no.  N202 166840 granted by the National Center for Science of Poland. This work has been supported by the RFBR projects no. 12-03-31203 and 13-03-00125.

\end{document}